\documentclass{jps-cp}
\usepackage{txfonts} %Please comment out this line unless the txfonts package is availabe in your LaTeX system.

\title{The Precision nEDM Measurement with UltraCold Neutrons at TRIUMF}

\author{Ryohei \textsc{Matsumiya}$^{1,2}$, Hiroaki \textsc{Akatsuka}$^{3}$, Chris \textsc{P. Bidinosti}$^{4}$, Charles \textsc{A. Davis}$^{1}$, Beatrice \textsc{Franke}$^{1,10}$, Derek \textsc{Fujimoto}$^{1}$, Michael \textsc{T. W. Gericke}$^{5}$, Pietro \textsc{Giampa}$^{6}$, Robert \textsc{Golub}$^{7}$, Sean \textsc{Hansen-Romu}$^{4, 5}$, Kichiji \textsc{Hatanaka}$^{2}$, Tomohiro \textsc{Hayamizu}$^{8}$, Takashi \textsc{Higuchi}$^{2}$, Go \textsc{Ichikawa}$^{9}$, Sohei \textsc{Imajo}$^{2}$, Blair \textsc{Jamieson}$^{4}$, Shinsuke \textsc{Kawasaki}$^{9}$, Masaaki \textsc{Kitaguchi}$^{3}$, Wolfgang \textsc{Klassen}$^{10}$, Emma \textsc{Klemets}$^{10}$, Akira \textsc{Konaka}$^{1}$, Elie \textsc{Korkmaz}$^{11}$, Ekaterina \textsc{Korobkina}$^{7}$, Florian \textsc{Kuchler}$^{12}$, Maedeh \textsc{Lavvaf}$^{5}$, Larry \textsc{Lee}$^{1,5}$, Thomas \textsc{Lindner}$^{1,4}$, Kirk \textsc{W. Madison}$^{10}$, Yasuhiro \textsc{Makida}$^{9}$, Russell \textsc{Mammei}$^{1,4}$, Juliette \textsc{Mammei}$^{5}$, Jeffery \textsc{W. Martin}$^{4}$, Mark \textsc{McCrea}$^{4}$, Eric \textsc{Miller}$^{10}$, Kenji \textsc{Mishima}$^{9}$, Takamasa \textsc{Momose}$^{10}$, Takahiro \textsc{Okamura}$^{9}$, Hooi Jin \textsc{Ong}$^{2,14}$, Ruediger \textsc{Picker}$^{1,13}$, William \textsc{D. Ramsay}$^{1}$, Wolfgang \textsc{Schreyer}$^{1}$, Hirohiko \textsc{M. Shimizu}$^{3,9}$, Steve \textsc{Sidhu}$^{1,13}$, Shawn \textsc{Stargardter}$^{4,5}$, Isao \textsc{Tanihata}$^{2,15}$, Sean \textsc{Vanbergen}$^{1,10}$, Willem \textsc{T. H. van Oers}$^{1,5}$ and Yutaka \textsc{Watanabe}$^{9}$ (TUCAN Collaboration)}

\inst{$^{1}$TRIUMF, 4004 Wesbrook Mall, Vancouver, BC V6T 2A3, Canada \\
$^{2}$RCNP, Osaka University, 10-1 Mihogaoka, Ibaraki, Osaka 567-0047, Japan \\
$^{3}$Nagoya University, Nagoya, Aichi 464-8501, Japan \\
$^{4}$University of Winnipeg, Winnipeg, MB R3B 2E9, Canada \\
$^{5}$University of Manitoba, Winnipeg, MB R3T 2N2, Canada \\
$^{6}$SNOLAB, Lively, ON P3Y 1N2, Canada \\
$^{7}$North Carolina State University, Raleigh, NC 27695, USA \\
$^{8}$RIKEN, Wako, Saitama 351-0198, Japan \\
$^{9}$High Energy Accelerator Research Organization (KEK), Tsukuba, Ibaraki 305-0801, Japan \\
$^{10}$The University of British Columbia, Vancouver, BC V6T 1Z4, Canada \\
$^{11}$University of Northern British Columbia, Prince George, BC V2N 4Z9, Canada \\
$^{12}$The Technical University of Munich, 80333 Munich, Germany \\
$^{13}$Simon Fraser University, Burnaby, BC V5A 1S6, Canada \\
$^{14}$Institute of Modern Physics, Chinese Academy of Sciences, Lanzhou, China, 730000 \\
$^{15}$Beihang University, Haidian District, Beijing, P.R. China, 100191}

\email{matsumiya@triumf.ca}

\recdate{February 15, 2022}

\abst{The TRIUMF Ultra-Cold Advanced Neutron (TUCAN) collaboration aims at a precision neutron electric dipole moment (nEDM) measurement with an uncertainty of 10$^{-27}$\,$e\cdot$cm, which is an order-of-magnitude better than the current nEDM upper limit and enables us to test Supersymmetry. To achieve this precision, we are developing a new high-intensity ultracold neutron (UCN) source using super-thermal UCN production in superfluid helium (He-II) and a nEDM spectrometer. The current development status of them is reported in this article.}

\kword{ultracold neutrons, UCN, electric dipole moment, EDM, He-II}

\begin{document}
\maketitle

\section{Introduction}

Ultracold neutrons (UCNs) are extremely slow neutrons with energies below $\sim$300\,neV and totally reflected on material surfaces due to their long wavelength. UCNs are very interesting probes for fundamental physics because they can be stored in a vessel, manipulated and studied for a time as long as their lifetime of almost 15 minutes. The measurement of the neutron electric dipole moment (nEDM) is one of such studies. The magnitude of nEDM predicted by the Standard Model of particle physics (SM) is on the order of 10$^{-31}$\,$e\cdot$cm, while Supersymmetry (SUSY), a popular beyond the standard model (BSM) physics scenario, predicts 10$^{-27}\sim10^{-28}\,e\cdot$cm. Therefore, a search of the nEDM gives access to testing the BSM.

The latest measured value of the nEDM upper limit is $1.8\times10^{-26}\,e\cdot$cm (90\% C.L.)\cite{nedm}. In that measurement, the Ramsey resonance technique is applied on polarized UCNs stored in a vessel called EDM cell\cite{ramsey}. The uncertainty of the current measurement is dominated by a statistical error, i.e. UCN density in their EDM cell. Thus, it is essential to increase UCN density to achieve a higher sensitivity nEDM measurement.

The TUCAN (TRIUMF Ultra Cold Advanced Neutron) collaboration is developing a new high-intensity UCN source and a nEDM spectrometer that allows to measure the nEDM on the order of $10^{-27}\,e\cdot$cm and test SUSY. In the following sections, the scheme of our UCN source and the development status of our equipment are explained.

\section{Development of the high-intensity super-thermal UCN source}

\subsection{Principles of super-thermal UCN production}

The phase space density of UCNs extracted from cold neutron sources is limited by Liouville’s theorem, resulting in small UCN densities in experimental vessels and less measurement precision. To overcome this limitation, super-thermal UCN production was suggested, which converts cold neutrons to UCNs using down-scattering by phonons in superfluid helium (He-II)\cite{golub}. In this method, efficient UCN production becomes possible by using the phase space volume of phonons in He-II for neutron cooling.

\begin{figure}[tbh]
    \begin{center}
        \includegraphics[width=0.8\textwidth]{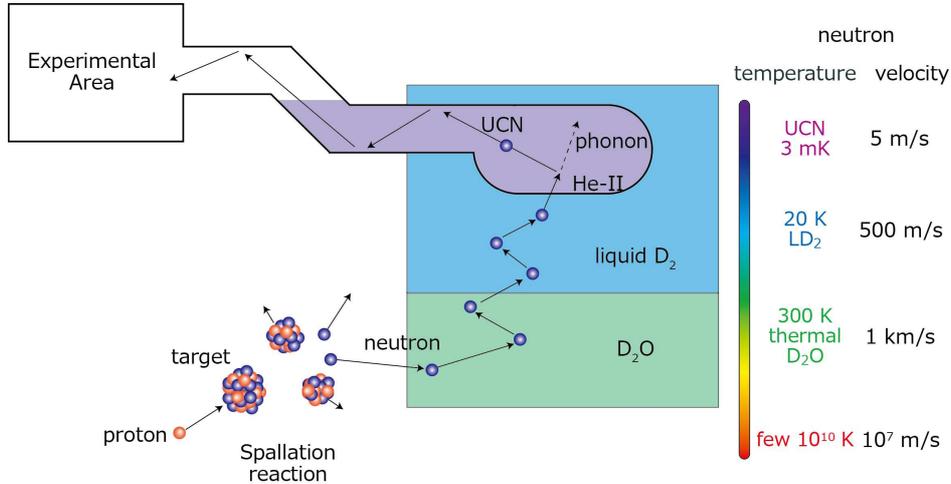}
        \caption{Scheme of the TUCAN super-thermal UCN source. Fast neutrons are produced by a spallation reaction on the tungsten target and cooled down to cold neutrons in moderators (green and blue). The cold neutrons are converted to UCNs in He-II (purple), then guided to the downstream experimental volume through UCN guides. Reproduced from\cite{SK2020}.}
        \label{superthermal}
    \end{center}
\end{figure}

We couple super-thermal UCN production with spallation neutron production driven by TRIUMF's main cyclotron. Figure\,\ref{superthermal} shows the scheme of our UCN source. Fast neutrons are produced through spallation on a tungsten target. These fast neutrons are moderated in 300\,K heavy water and liquid deuterium (LD$_2$) down to cold neutrons. The cold neutrons are converted to UCNs in He-II, then transported to an experimental vessel.
The produced UCNs can be lost due to up-scattering by phonons in He-II. As a result, the actual UCN lifetime becomes shorter than the neutron $\beta$-decay lifetime. To suppress this loss, the He-II needs to be cooled down to $\sim$1\,K.

\subsection{Prototype UCN source}

\begin{figure}[tbh]
%\hspace{0.025\linewidth}
%\begin{minipage}{0.45\linewidth}
\centering
        \includegraphics[width=0.95\linewidth]{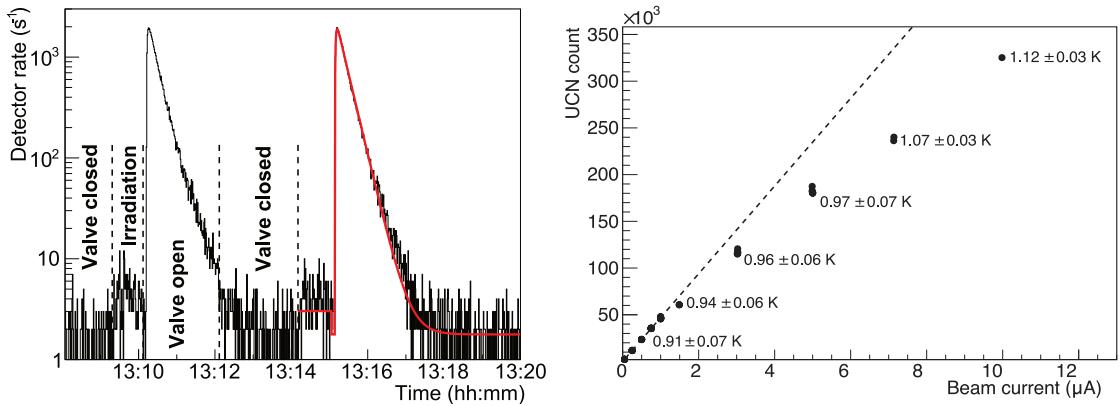}
        \caption{Results obtained during the 2017 beamtime at TRIUMF. Reproduced from\cite{tucanucn2019}. (Left) - Typical UCN counts observed by the UCN detector attached to the exit of the UCN source. (Right) - Number of UCN counts extracted from the source with different proton beam currents. The labels indicate the peak He-II temperatures reached during irradiation.}
        \label{FigureProtoUCNsource}
%\end{minipage}%
%\hspace{0.05\linewidth}
%\begin{minipage}{0.45\linewidth}
%\centering
%        \includegraphics[width=0.95\linewidth]{Nr_UCN_current.eps}
%        \caption{Number of extracted UCNs as a function of beam current}
%        \label{UCNcounts_beamcurremt}
%\end{minipage}
\end{figure}

In order to test the feasibility of super-thermal UCN production, the development of a prototype UCN source was started at KEK in the 1990s. It was installed and successfully operated at the Research Center for Nuclear Physics (RCNP), Japan\cite{masuda2002}. After 10 years of development, a UCN density of 26\,UCN/cm$^{3}$ was achieved\cite{masuda2012}. This UCN source was moved to TRIUMF in 2017\cite{tucanucn2019}. Figure\,\ref{FigureProtoUCNsource} (Left) shows typical UCN counts observed by the UCN detector attached to the exit of the UCN source in the 2017 TRIUMF beamtime. During that beamtime, the proton beam current was increased up to 10\,$\mu$A and the UCN counts were measured as a function of beam current. The result is shown in Fig.\,\ref{FigureProtoUCNsource} (Right). The UCN yield was proportional to the current below 1\,$\mu$A, while the yield dropped at higher currents due to the rise of He-II temperature induced by the increased heat load stemming from $\gamma$- and $\beta$-radiation from the spallation target and activated UCN production bottle, respectively. This is expected as the prototype source cryostat was developed and optimized to deliver cooling power corresponding to the heat load caused by no more than 1\,$\mu$A.

\subsection{Scheme of the TUCAN source}

\begin{figure}[tbh]
    \begin{center}
        \includegraphics[width=1.0\textwidth]{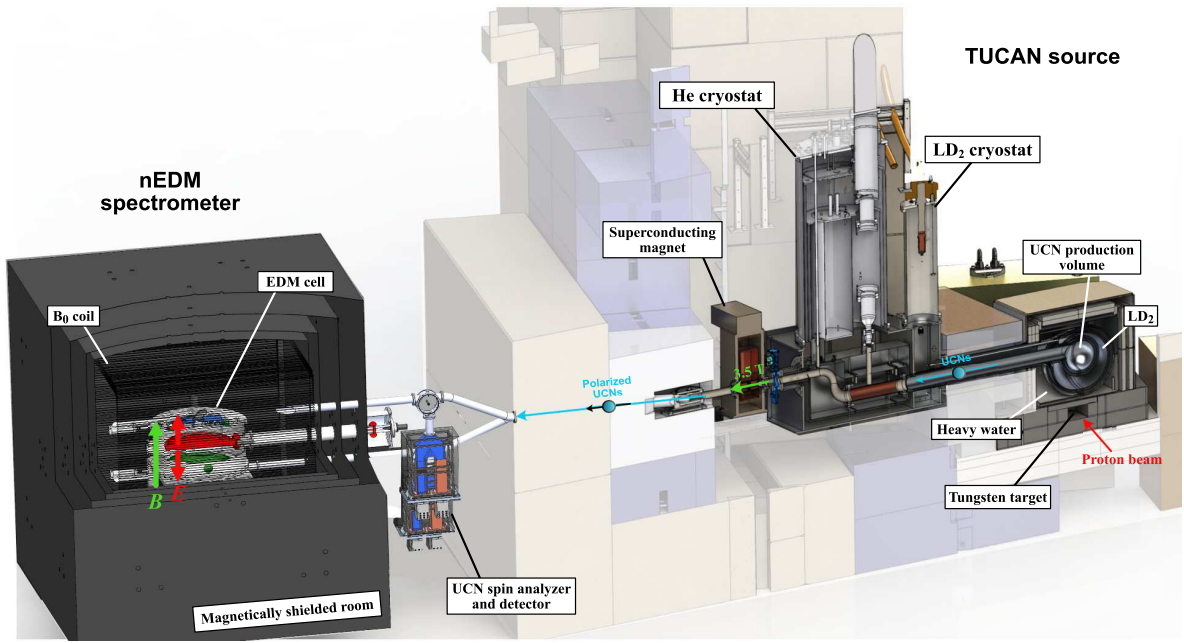}
        \caption{Overview of the TUCAN source and the nEDM spectrometer. Fast neutrons produced by spallation onto a tungsten target are moderated to cold neutron energies, then converted to UCNs in the produciton volume filled with He-II. The UCNs are extracted from the source and polarized by a 3.5\,T magnetic field of a superconducting magnet, then guided into the EDM cell. After spin manipulation by the Ramsey resonance technique, the UCNs are extracted from the cell for spin analysis and detection.}
        \label{TUCANsource}
    \end{center}
\end{figure}

Following the successful operation of the prototype UCN source, the TUCAN collaboration started the development of a new high-intensity UCN source, the TUCAN source (Fig.\,\ref{TUCANsource}). This new UCN source will allow for UCN production with proton beam currents up to 40\,$\mu$A, while the prototype UCN source was usually operated at 1\,$\mu$A. In addition to the beam current upgrade, the cold moderator will be upgraded from solid heavy water to LD$_2$. The UCN production volume is also increased from 8\,L to 27\,L. With these improvements, the UCN production rate is expected to be 1.4 to $1.6 \times 10^7$\,UCN/s, which is about 1000 times as large as in the prototype UCN source\cite{moderator2020}.

Figure\,\ref{TUCANsource} shows the nEDM spectrometer connected to the downstream side of the TUCAN source. The extracted UCNs are polarized by passing through a 3.5\,T magnetic field generated by a superconducting magnet. These polarized UCNs are guided into the EDM cell, where a static magnetic field $B$ and electric field $E$ are applied. The EDM cell is placed in a magnetically shielded room (MSR), where the internal magnetic field is precisely monitored and controlled. After applying Ramsey's method of separated oscillatory fields to the UCN stored in the EDM cell, UCNs are extracted from the cell and their spin direction is analyzed. The statistical uncertainty of nEDM measurement is written as: $\delta d_n \sim \hbar/2ET \alpha \sqrt{N}$, where $T$ is the free precession time, $E$ is the strength of the electric field, $\alpha$ is the visibility of the Ramsey fringe and $N$ is the number of detected UCNs. The expected UCN density in the EDM cell with our TUCAN source is $\sim$250\,UCN/cm$^{3}$, or $N \sim$10$^{6}$ per cycle. Assuming $T \sim 100$\,sec, $E \sim $10\,kV/cm and $\alpha \sim 1$, $\delta d_{n} \sim 10^{-25}\,e\cdot$cm per each Ramsey cycle is obtained. $10^{-27}\,e\cdot$cm can be reached within $\sim$ 400 days of proton beam operation.

\section{Recent developments of the TUCAN source and nEDM spectrometer}

\subsection{TUCAN source - cooling test of the He cryostat}

\begin{figure}[tbh]
%\begin{minipage}{0.45\linewidth}
    \begin{center}
        \includegraphics[width=1.0\linewidth]{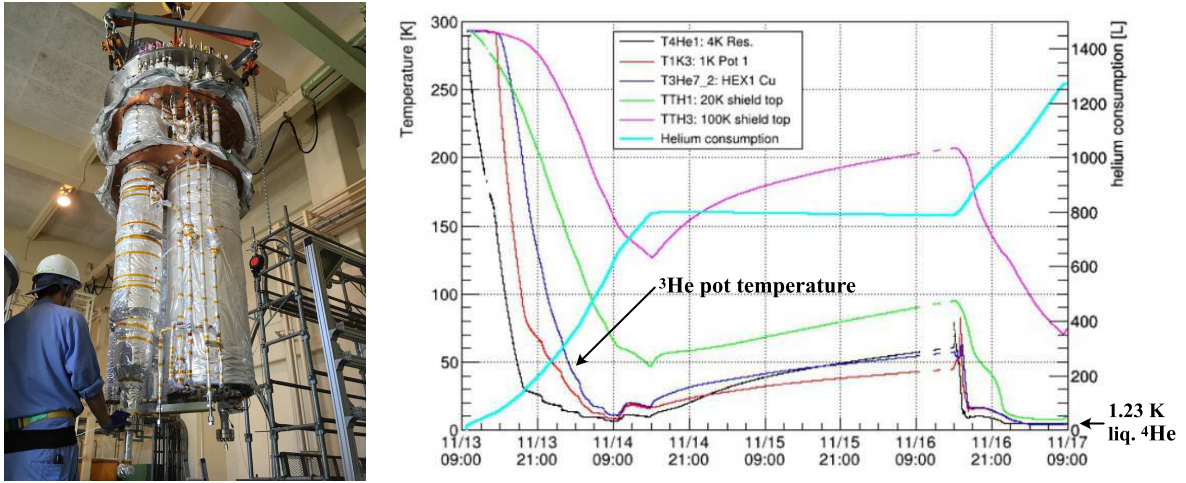}
        \caption{(Left) - Photograph of the He cryostat, taken during the 2020 cooling test at KEK. (Right) - Time evolution of temperatures inside the He cryostat during the cooling test. The blue line shows the $^{3}$He pot temperature. Liquid $^{4}$He was condensed in the $^{3}$He pot and cooled down to 1.23\,K.}
        \label{He_cryostat}
    \end{center}
%\end{minipage}
\end{figure}

We are planning to operate the TUCAN source with a 20\,kW proton beam. Since the heat load is expected to be $\sim$10\,W during beam irradiation, it is essential to strengthen the cooling power of the UCN source. The helium (He) cryostat (Fig.\ref{He_cryostat} (Left)), one of the crucial subsystems of the TUCAN source, was designed to achieve this cooling power\cite{SK2020, Okamura2020}. It cools liquid $^{3}$He down to 0.8\,K by pumping with 8,800\,m$^{3}$/h vacuum pumps, then He-II is condensed in the UCN production volume and cooled down to $\sim$1\,K via a heat exchanger attached between the $^{3}$He pot and the UCN volume. The construction of the He cryostat was started at KEK in 2019. In 2020, a cooling test of the He cryostat was done and liquid $^{4}$He was successfully condensed in the $^{3}$He pot. The liquid $^{4}$He was cooled down to 1.23\,K (Fig.\ref{He_cryostat} (Right)), corresponding to 0.65\,K for liquid $^{3}$He at the same saturation vapor pressure. The He cryostat was shipped to TRIUMF in 2021 and re-assembly and initial testing has begun.

In parallel, developments of the other subsystems are underway. UCN guides with various Nickel-Phosphorus (NiP) platings were tested using either the prototype UCN source or the UCN source at J-PARC. UCN storage experiments of the UCN production volume were performed at Los Alamos National Laboratory (LANL) in 2021. The LD$_2$ cryostat, the $^{3}$He/$^{4}$He gas handling systems and the liquid helium transfer line are being designed and constructed at TRIUMF. With these systems, we are aiming at UCN production in 2022/2023.

\subsection{nEDM spectrometer}

The static magnetic field $B$ used for the nEDM measurement will be $\sim$1\,$\mu$T, which is 50 times smaller than Earth’s field. Therefore, our EDM cell is placed inside an MSR consisting of four layers of mu-metal. The design shielding factor is on the order of 10$^{5}$. In the $1\times1\times1$\,m$^{3}$ central volume of the MSR, the residual magnetic field will be $<$1\,nT and the field gradient will be $<$ 100\,pT/m. The installation of the MSR is planned in 2022. Prior to designing the MSR, a three-dimensional magnetic field mapping in the location of the MSR was performed. Based on the mapping data, a set of compensation coils which reduces magnetic flux in the mu-metal layers and ensures the performance of the MSR are being designed\cite{higuchi2021}. The field fluctuation inside the EDM cell and its gradient will be measured by the use of a $^{199}$Hg co-magnetometer\cite{199hg} and Cs magnetometers\cite{csmagneto}. A prototype EDM cell and valve were constructed at TRIUMF. We’re planning to test them at J-PARC in spring 2022. The UCN spin analyzer is under development. It analyzes UCN spins by measuring UCN transmission through a thin magnetized iron foil.%The iron foil was developed using the ion-beam sputtering (IBS) facility at the Institute for Integrated Radiation and Nuclear Science, Kyoto University (KURNS), then tested using the cold neutron beam at J-PARC.
The foil was already tested using a cold neutron beam at J-PARC, and an additional test with UCNs is also planned in spring 2022\cite{higuchi2021,akatsuka2021}.

\section{Conclusion}

nEDM is a critical probe to test BSM theories like SUSY. The TUCAN collaboration aims to measure nEDM with a sensitivity of $10^{-27}\,e\cdot$cm, which is better than the current limit by a factor of 10. To overcome the statistical limitation, the TUCAN collaboration is developing a high-intensity super-thermal UCN source, based on the results of a prototype UCN source development and its operation.

Recent progress on developments of the new UCN source (TUCAN source) and the nEDM spectrometer are reported. We are planning to install the TUCAN source and produce UCN in 2022/2023 and start commissioning of the nEDM spectrometer in 2023. 

\section*{Acknowledgment}

The authors would like to express their sincere gratitude to C. Marshall, T. Hessels, J. Chak, S. Horn and D. Rompen (TRIUMF) for their engineering support.

The work of the TUCAN collaboration is financially supported by the Canada Foundation for Innovation (CFI), the Canada Research Chairs program, the Natural Sciences and Engineering Research Council of Canada (NSERC), B.C. Knowledge Development Fund (BCKDF), Research Manitoba, the Japan Society for the Promotion of Science (JSPS), Yamada Science Foundation, and the RCNP Collaboration Research network (COREnet) program.

\end{document}